# Zero-waste manufacturing of ophthalmic lenses by direct Fluidic Shaping in arbitrary domains

Yotam Katzman[1], Mor Elgarisi[1], Amos A. Hari[1], Jonathan Ericson[1], Omer Luria[1], Valeri Frumkin[1,2], Moran Bercovici[1,*]

[1] Faculty of Mechanical Engineering, Technion – Israel Institute of Technology, Haifa, Israel

[2] Current affiliation: Department of Mechanical Engineering, Boston University, MA, USA

* Corresponding author: mberco@technion.ac.il

## ABSTRACT

The conventional manufacturing of ophthalmic lenses is an inefficient subtractive process where up to 97% of the material is discarded through grinding, polishing, and edging. Fluidic Shaping has emerged as a powerful alternative, utilizing surface tension to form optical-quality surfaces. While the approach enabled the creation of ophthalmic lenses without grinding or polishing, it was limited to lenses with a circular or elliptical footprint and still required the wasteful edging process to fit the lenses into the eyewear rims. Here, the *Cookie Cutter* algorithm is introduced, generalizing the Fluidic Shaping approach to be applicable to arbitrary domains, thus eliminating all subtractive processes. This mathematical framework calculates the unique varying edge-height required for a boundary frame, allowing a liquid polymer to naturally settle into a target spherocylindrical prescription within an arbitrary rim footprint. By utilizing neutral buoyancy to negate gravity, the liquid polymer is shaped solely by surface tension and subsequently cured, resulting in a lens that fits directly into commercial eyewear rims without any mechanical post-processing. The method is validated experimentally, demonstrating the fabrication of lenses compatible with standard eyewear rims. This approach represents a complete additive manufacturing solution, enabling end-to-end zero-waste production of prescription eyeglasses.

## INTRODUCTION

The global eyewear industry produces around 2-3 million lenses each day in order to meet the high demand for optical components, primarily in the form of prescription lenses,[1–4] and sunglasses[3,5]. The conventional methods for manufacturing those lenses rely extensively on machining processes such as griding, polishing, and edging, which are time and energy consuming, and generate more than 50,000 tons of polymeric waste each year (see SI.1 for details).[3,6–8]

The conventional lens manufacturing process begins with the synthesis of raw materials, such as CR-39, polycarbonate, or high-index polymers.[1,2,9–13] These materials are produced by few central manufacturers (e.g. PPG (USA)[14], Mitsui (Japan), Covestro (Germany)), and are then shipped under controlled thermal and humidity conditions to casting facilities in other regions in the world, primarily in China, Thailand and India. At these facilities, the polymers are casted or injection molded,[9,13–22] to form either complete simple single-prescription lenses or semi-finished blanks (SFBs)[13,23,24] - large blocks of material that serve as the basis for subsequent machining. Their cost depends mainly on the refractive index of the material and ranges from $0.50 to $15 per blank. The SFB are shipped to finishing labs all over the world where they are surfaced according to the customer's prescription and eyewear rim.

The surfacing stage – in which the desired optical surface is produced – begins with CNC grinding machines followed by precision polishing,[23,25–27] to achieve the desired curvature and optical quality. The final surfacing step is edging[13,23,25,28], where the lens is cut to fit the exact shape and size of the eyewear rim using high-speed rotary cutters. In the surfacing process 90-97% of the material is discarded - a typical semi-finished blank weighs 50-70 grams (see table S.1), and results in a final lens of 2-5 grams.[13] i.e. most of the material, which has been shipped twice around the globe, is eventually discarded at the final step of manufacturing. Thus, many optical companies invest significant efforts to develop new methods to reduce material waste. For instance, Essilor-Luxottica developed an optimization algorithm allowing for better matching between lens blank size and frame dimensions.[24,29] This approach reduces unnecessary material removal during surfacing and can lower waste by up to 30–40% though it does not eliminate the inherent wastefulness of the process. Furthermore, facilities must maintain a large inventory of semi-finished lenses of different initial diameters and curvatures, to meet the wide range of rims and prescriptions.[23]

Fluidic Shaping is a recently developed alternative approach for optical manufacturing that is based on manipulation of liquid polymers rather than subtractive machining.[30,31] In Fluidic Shaping an optical polymer is injected into a bounding frame contained in an immersion density-matched liquid, allowing buoyancy forces to negate the effect of gravity. Under these conditions, surface tension dominates and drives the liquid interface into a desired minimum energy state resulting in a liquid lens. The liquid lens can then be solidified within minutes by UV or thermal curing. The optical properties of the lens are dictated by the volume of the liquid and the geometry of the frame, and Elgarisi et al. showed that by using simple elliptical frames with varying aspect ratios, any sphero-cylindrical prescription can be achieved.[32] The resulting lenses fabricated in this method do not require grinding nor polishing processes and yield optical power deviations of less than 0.125 diopters, and surface roughness of less than 1 nm RMS – meeting, and in many cases surpassing, the industrial standards.

To date, Fluidic Shaping was shown to be an alternative to machining and polishing. However, subtractive processes and material waste could not be entirely avoided in producing the final lenses, as edging was still necessary in order to fit the lenses into the eyewear rims. Here, we introduce the '*Cookie Cutter*' algorithm – a simple mathematical approach that identifies the unique boundary conditions required to fabricate a desired prescription lens that has the footprint of the customer's eyewear rim. By implementing Fluidic Shaping on frames designed by the Cookie Cutter algorithm, we show the feasibility for end-to-end fabrication of ophthalmic lenses with any spherical and cylindrical prescription on any eyewear rim shape, in which all subtractive processes and their associated material waste are entirely eliminated.

**THEORY AND PRINCIPLE OF THE METHOD**

We here derive the mathematical formulation that uniquely defines the height variation along a frame with an arbitrary footprint, such that upon injection of polymer into it, the obtained lens will achieve the desired optical prescription. As illustrated in Figure 1, the result can be interpreted as the action of a cookie cutter, and we thus term it 'the Cookie Cutter algorithm'.

The Fluidic Shaping approach is based on injection of a liquid polymer into a bounding frame submerged with an immiscible immersion liquid of equal density. The polymer settles into a minimum-energy state with its free interfaces, $h^{(t)}, h^{(b)}$ serving as optical surfaces (where the t and b superscripts denote the top and bottom surfaces, respectively), which satisfy the partial differential equation (PDE),

$$\begin{cases} \dfrac{2h_{xy}^{(t)}h_x^{(t)}h_y^{(t)} - h_{xx}^{(t)}\left(1+\left(h_y^{(t)}\right)^2\right) - h_{yy}^{(t)}\left(1+\left(h_x^{(t)}\right)^2\right)}{\left(1+\left(h_x^{(t)}\right)^2+\left(h_y^{(t)}\right)^2\right)^{\frac{3}{2}}} = \lambda^{(t)} \\ \dfrac{2h_{xy}^{(b)}h_x^{(b)}h_y^{(b)} - h_{xx}^{(b)}\left(1+\left(h_y^{(b)}\right)^2\right) - h_{yy}^{(b)}\left(1+\left(h_x^{(b)}\right)^2\right)}{\left(1+\left(h_x^{(b)}\right)^2+\left(h_y^{(b)}\right)^2\right)^{\frac{3}{2}}} = \lambda^{(b)} \end{cases}, \quad (1)$$

where the subscripts $x$ and $y$ indicate derivatives with respect to the $x$ and $y$ coordinates, and $\lambda^{(b)}$ and $\lambda^{(t)}$ are Lagrange multipliers which correspond to volumetric constrains. For the case of a symmetric (bi-convex or bi-concave) lens, $\lambda^{(b)} = -\lambda^{(t)}$, and the constraint is simply that the volume contained between the surfaces is the injected polymer volume, $V_{theoretical\ lens} = \iint h^{(t)} - h^{(b)} dxdy$. The polymer can then be cured, resulting in a solid lens.

For the case in which $\Gamma(x,y)$ is an elliptical contour given by $\dfrac{x^2}{a^2} + \dfrac{y^2}{b^2} = 1$, and the frame's height is uniform, Elgarisi *et al.* showed that the linear approximation to Equation 1 yields a set of surfaces corresponding to spherical and cylindrical corrections.[32] However, lenses produced in this way still require edging in order to fit into desired rims. A natural question thus arises – can such optical prescriptions be obtained on non-elliptical frame shapes, thus eliminating also the edging process? Unfortunately, direct analytical solutions of the partial differential equations are not feasible on arbitrary domains. Nonetheless, the solution can be obtained using a simple thought experiment, as illustrated in Figure 1: suppose that we have an optical design for a lens that could be obtained using the Fluidic Shaping method on an elliptical frame that encloses a domain $\Omega$. The surfaces of this theoretical lens satisfy the PDE, Equation 1 over the domain, with pinning boundary conditions on its edge. The real lens we wish to create should fit into an eyewear rim whose planar projection is the contour $\tilde{\Gamma}$ (depicted by the purple contour in Figure 1a and 1b) that encompasses a domain $\omega$. As long as $\omega$ is fully contained in $\Omega$, i.e. $\omega \subset \Omega$, we can define new functions, $v^{(t)}(x,y), v^{(b)}(x,y)$ such that $v^{(t)}(x,y) \equiv h^{(t)}(x,y)$ and $v^{(b)}(x,y) \equiv h^{(b)}(x,y)$ inside $\omega$. By definition, the surfaces $v^{(t)}$ and $v^{(b)}$ have the same topography as $h^{(t)}$ and $h^{(b)}$, and they satisfy Equation 1 subjected to the boundary conditions

$$v^{(t)}(\tilde{\Gamma}) = h^{(t)}(\tilde{\Gamma}) \qquad (2a)$$

$$v^{(b)}(\tilde{\Gamma}) = h^{(b)}(\tilde{\Gamma}) \qquad (2b)$$

with a polymer volume of

$$V_{lens} = \iint_\omega \left(h^{(t)} - h^{(b)}\right) dxdy = \iint_\omega \left(v^{(t)} - v^{(b)}\right) dxdy \qquad (3)$$

Thus, a frame with an arbitrary footprint shape and non-uniform height along the boundaries defined by Equations 2a-b, into which a volume defined by Equation 3 is injected, would recreate the original optical surfaces within the subdomain $\omega$. Figures 1c.i-ii illustrate the defined frame design and the required calculated injection volume, respectively.

A practical implementation of this algorithm is depicted in Figure 1a-c. Consider the intersection of an infinite cylinder, formed by the extrusion of the domain $\omega$ in a direction perpendicular to its plane, with a theoretical lens designed on an elliptical domain. The resulting three-dimensional shape is the desired final lens that fits the eyewear rim, with its top and bottom surfaces coinciding with the optical surfaces of the original lens. Its perimeter surface defines a Fluidic Shaping frame (Figure 1c.i) with variable height, and its volume corresponds to $V_{lens}$ (Figure 1c.ii) – the exact volume that should be injected into the frame to form the desired lens.

Figures 1d-f illustrate the lens fabrication process. We 3D print the designed frame and insert it into a container filled with an immersion liquid of density $\rho_{im}$ that is set to match the density of the optical polymer ($\rho_{im} = \rho_{lens}$), such that gravity and buoyancy forces cancel each other. We inject the exact nominal volume ($V_{lens}$) into the frame, ensuring that the polymer wets the entire inner surface of the frame and is pinned at its edges. The optical polymer is immiscible with the immersion liquid, and in this effective weightlessness environment, the dominant force is surface tension, which derives the liquid volume into its minimum energy state. Within seconds, the liquid polymer settles into the desired shape, resulting in a liquid lens. The lens is then cured under UV for several minutes, which results in a solid lens with the designed optical properties. Finally, we remove the solid lens from the fabrication system and insert it directly into the eyewear rim, without any post-processing or material waste.

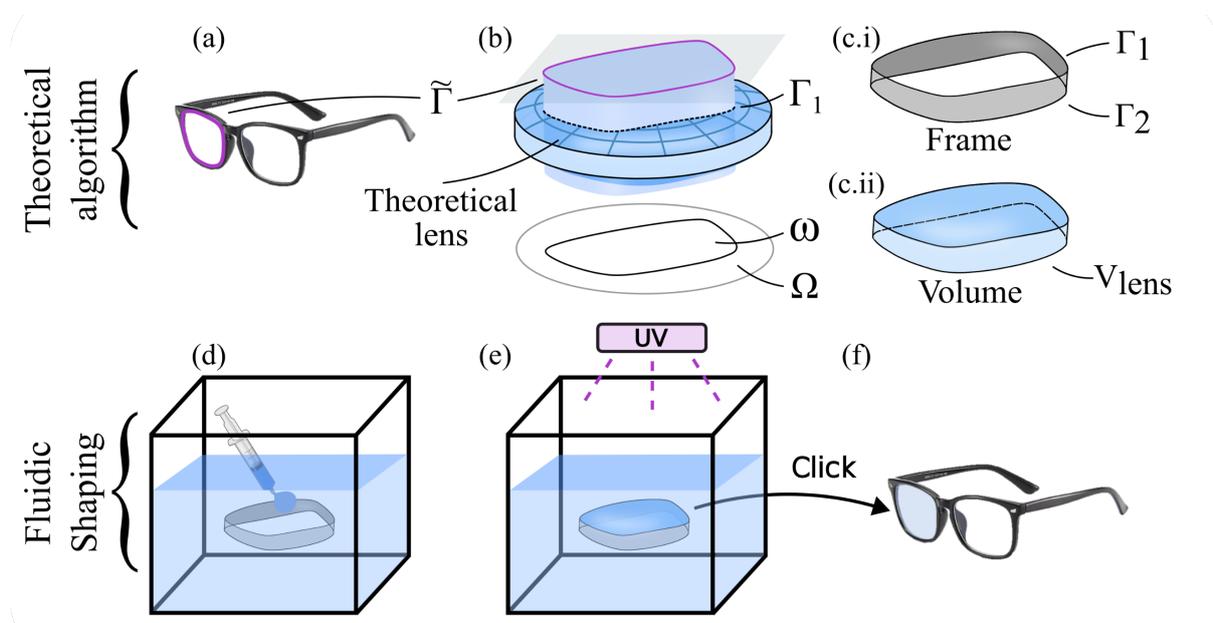

*Figure 1 – The Cookie Cutter algorithm and fabrication process via Fluidic Shaping. First row: the theoretical Cookie Cutter algorithm. (a) We obtain a 2D contour that has the exact footprint of the desired eyewear rim, $\tilde{\Gamma}$. (b) A theoretical process of intersection between the rim-shaped contour and a theoretical lens that was designed with specific optical properties. This intersection process results in (c.i) a new varying edge-height frame for later fabrication and (c.ii) the precise volume needed to be injected into the new frame in order to achieve the desired lens and optical properties. Second row: the Fluidic Shaping lens fabrication process. (d) We 3D-printed the frame and submerged into a container filled with an immersion liquid. (e) We injected the precise volume of the optical polymer into the frame, results in a liquid Fluidic Shaping lens that will be cured under UV for several minutes. (f) The solid lens has the exact spherical and cylindrical powers as the theoretical lens and has the same footprint as the eyewear rim. Therefore, can be directly inserted into the eyewear rim without any further processing or material waste.*

# EXPRIMENTAL RESULTS

## Proof of concept

Figure 2 presents a qualitative assessment for the performance of a lens produced using the Cookie Cutter algorithm applied to a commercial Oakley frame. For reference, we also present the results obtained from a frame with an identical footprint, but with a 'naïve' uniform height, i.e. 'flat'. Both frames were mounted side by side in a container filled with immersion liquid. As there is no nominal volume that will result in the desired lens for the flat frame, we injected the nominal volume of the varying edge-height frame into each of the frames. We designed the "Cookie Cutter" frame based on a circular bi-convex Fluidic Shaping lens with a spherical diopter of $P = 7$ $m^1$ and no astigmatism ($C = 0$ $m^1$). We purposely designed this lens with a strong spherical power because its effective diopter under the immersion liquid is significantly smaller, since the light now passes through two media with a much smaller difference in index of refraction (in air: n=1.525 for the solid polymer vs n=1 for air; in water: n<1.525 for the liquid polymer vs n=1.36 for the water. See SI.2 for details). Figure 2 shows that, as expected, the flat frame yields a highly distorted lens because its boundary conditions are incompatible with a spherical surface, for any injected volume. In contrast, by injecting the exact nominal volume into the Cookie Cutter frame, we obtain a lens that delivers an undistorted uniformly magnified image corresponding to the effective diopter, indicating that the varying edge height frame preserves the intended pure spherical power. Figure S.3 in the Supplementary Information presents another example of the same frame but with a smaller injected volume that results in a lens with similar radii of curvature to those of the ideal lens. As expected, this lens also results in significant optical distortions.

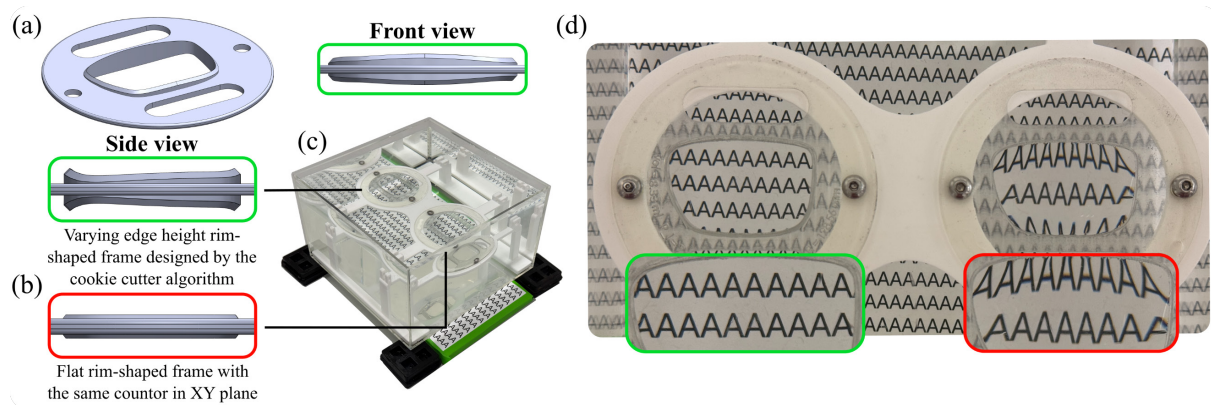

*Figure 2* – Demonstration and qualitative assessment of a lens produced by the Cookie Cutter algorithm applied to a commercial Oakley rim, and comparison to a lens produced on a frame with an identical footprint but uniform height. (a-b) CAD isometric and side views of two frames that have the same footprint but differ in their edge height. The lower frame has the same height along its entire contour (i.e. is flat), whereas the upper frame has a varying edge-height as designed by the Cookie Cutter algorithm. (c) Image of the fabrication set-up consisting of a container filled with immersion liquid, a holder supporting the two frames, and a patterned slider located underneath the container at a distance of 7.5 cm from the frame's plane. (d) Both frames were filled with the exact same nominal volume corresponding to the required volume of the varying edge-height frame. The flat frame (right, red) exhibits noticeable astigmatism and distortions, whereas the varying edge-height frame (left, green) achieves the desired pure spherical correction which appears in the image as an undistorted uniform magnification of the characters.

This qualitative experiment clearly validates our theory under nominal conditions. However, since for each frame designed by the Cookie Cutter algorithm there exists a unique injected volume (defined by Equation 3) that would yield the desired optical properties, it is important to quantify the sensitivity of these properties to the injection volume.

**Experimental setup**

To measure the effect of injected volume on the optical power of a lens, and to decouple it from other potential sources of deviation (e.g. shrinkage due to polymerization, as discussed in the following sections), we designed an experimental setup that allows us to measure the optical power of the lens while it is still in liquid form. At the heart of the setup is a Moiré deflectometer (Mapper, Rotlex, Israel) that provides a complete map of the lens' spherical and cylindrical powers based on a single image. However, this device is designed to measure solid optical components that can be placed into its measurement chamber. To measure the lenses while still in liquid form, we designed a sealed capsule that houses the liquid lens, and which fits in the Mapper's chamber. As illustrated in Figure 3a, the capsule consists of an SLA 3D printed chassis, a stage for positioning the frame, and two rubber o-rings which seal the capsule against two transparent acrylic glass plates through which the optical reading is performed. The fact that the liquid lens is measured in a liquid environment with an index of refraction different than that of air, as well as the existence of the two acrylic plates, creates an inherent deviation in the optical reading. To compensate for that, we measured several solid lenses both outside the capsule (in air) and inside the capsule (in the immersion liquid) and used those measurements to construct a transformation function between the two conditions. Additional information on this calibration process is provided in SI.4. Accordingly, all optical power measurements we present hereafter for the liquid lenses are provided as if they were solid lenses (with the same refractive index) used in air, although the measurements were in practice held in the capsule while they were in liquid form.

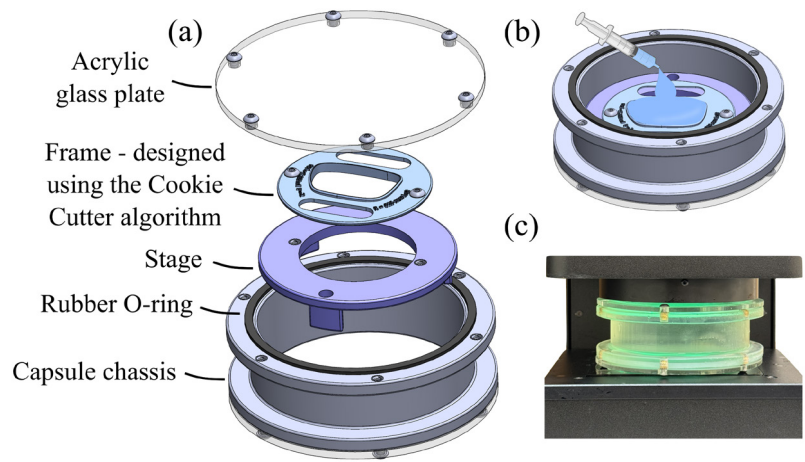

*Figure 3* – *Experimental setup for measuring the optical power of lenses while still in liquid form. (a-b) Exploded and assembled CAD illustration of the capsule designed for liquid lenses measurement. (c) Image of the capsule assembled on the Rotlex-Mapper device. The dimensions of the capsule are designed to fit into the maximum gap allowed by the mapper.*

**Spherical power sensitivity to deviations in the injection volume**

For a circular frame, theory predicts a linear change in lens power with variation in the injected volume.[32] We can expect other frame shapes to exhibit similar dependence of power on volume deviation, though with different sensitivity that depends on their geometry. Figure 4 presents both experimental and computational results for the dependence of lens power on volume deviations for various frame geometries. The experimental results are obtained through measurement of the lens spherical power in the capsule and its transformation to equivalent power in air, as described in SI.5. The computational results are obtained from our numerical simulation,[33] which solves the shape and curvature (that is uniform in these cases), then is multiplied by (n-1) to yield the spherical power. We investigate three representative frames on which we design lenses using the Cookie Cutter algorithm – an elliptical frame, a small children's eyewear rim based on a commercially available Oakley frame (Oakley Marshal Xs OY 8005), and a 25% scaled-up version of that same Oakley design, representing a typical large adult-sized eyewear rim. Each data point represents an experimental measurement of the average diopter over a region of 3×3 mm at the center of the lens (the complete data for these experiments is provided in SI.6 Table S.4). The linear lines correspond to the numerical solution for each frame geometry. There is a very good agreement between the experiments and the simulations, which validates the numerical solution. We can therefore rely on the numerical model to compute the volume accuracy required to achieve the industry standard of ±0.125 diopter, for any frame shape. To cover the entire range of lens sizes, we simulated the spherical power sensitivity for another frame, based on the commercial Oakley Slender, that serves as an extra-large adult's lens. Using those simulations, we find that an injection accuracy in the range of ±23–125 μL is required to meet the industry standard for all frames. In SI.7 we present a similar analysis for the cylindrical power sensitivity. We obtain that the spherical power is more sensitive to volume deviations and therefore dictates the injection accuracy requirements.

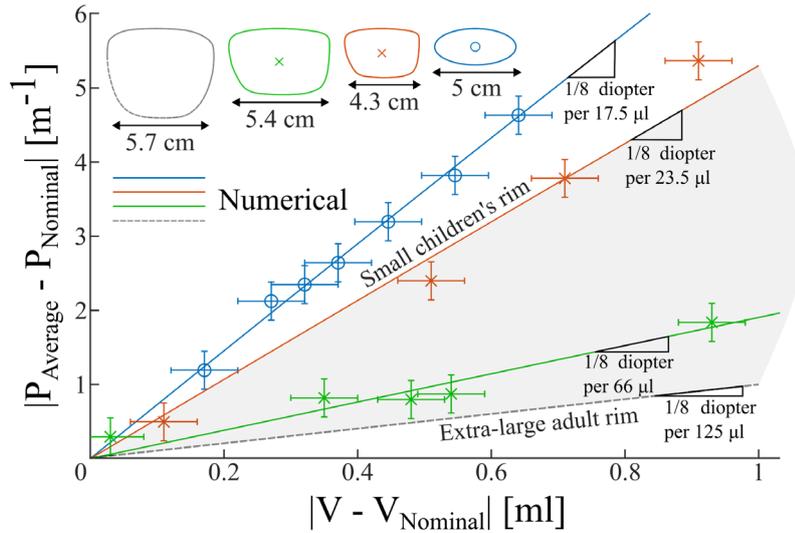

***Figure 4*** – *Experimental and numerical results of spherical power sensitivity to volume variations, for liquid lenses formed in three representative frame designs. The plot shows the absolute deviation in spherical power as a function of the absolute deviation in injected volume from its nominal value. Experimental data presents the average measured diopter values within a 3×3 mm square at the center of the lens for each tested injection volume. Horizontal error bars represent the uncertainty in injection volume (±50 µL), whereas vertical error bars show two standard deviations from the mean over the measured area (±0.2562 m$^{-1}$). The linear lines correspond to numerical solution for each frame geometry. The three representative frames include: (1) a simple elliptical frame (blue, o), (2) a small children's frame based on an Oakley eyewear rim (Oakley Marshal Xs OY 8005 Children's Glasses) (red, ×), and (3) a 25% scaled-up version of that same Oakley design, representing a large adult-sized eyewear rim (green, ×). The dashed line presents the numerical solution obtained for an extra large adults' frame (Oakley Slender). The grey shaded area between the gray and the red lines thus reflects the range of typical frame sizes. As derived from the slopes of the dashed lines, an injection accuracy in the range of ±23 (for very small frames) to 125 µL (for very large frames) is required to remain within the industry standard deviation of ±0.125 diopter.*

**An end-to-end demonstration of Cookie-Cutter based eyeglasses fabrication**

Figure 5 presents a complete set of eyewear glasses fabricated solely using Fluidic Shaping and the Cookie Cutter algorithm, without machining, edging or any post processing. We designed the nominal lens with a spherical power of $P = 4$ m$^{-1}$ and no cylindrical power, and applied the Cookie Cutter algorithm using the footprint of an Oakley Marshal Xs OY 8005 rim. We 3D-printed the computed frame on a Phrozen SLA printer (Phrozen Sonic Mighty Revo 14K using hyperfine resin), giving particular attention to orientation and supports in order to optimize the quality of the frame edges.

The theory described in section "Theory and principle of the method", refers to the shape of liquid polymer prior to polymerization. Naturally, any polymer experiences volumetric shrinkage during polymerization, which causes some deviation from the ideal surfaces and thus from the designed prescription. In Figures 2 and 4 we used a photopolymer - TJ-3704A, a commercial acrylate-based resin also sold under the brand name 'Vida Rosa' (Dongguan Tianxingjian Electronic Technology Co., Ltd, Guangdong, China) that has a volumetric shrinkage of 4.6%[34] (Reference SI – table 6) and polymerizes under UV in minutes, thus enabling a large number of experiments to be conducted in sequence. In Figure 5 we transitioned to RTV-601, a two-component optical silicone polymer that requires several hours to cure, but whose volumetric shrinkage is lower (its precise volumetric shrinkage is not provided by the manufacturer, but based on the datasheets provided by the manufacturer,[35] its volumetric shrinkage is estimated to be less than 1%), and thus yields smaller deformations upon polymerization. Because the refractive index of RTV-601 is different than that of Vida Rosa, we re-calibrated the capsule and derived a new transfer function for the RTV-601, as detailed in SI.4. We also derived the RTV-601 diopter sensitivity to volume for the Oakley frame, based on the simulations shown in Figure 4d and in SI.8.

To fabricate the lens, we injected an initial volume close to the nominal one, but without particular attention to its accuracy. We then measured the diopter of the liquid lens under the mapper, transformed it through the calibration transfer function, and used the sensitivity curve to deduce the amount of liquid that should be added or removed to reach the desired optical prescription. We repeated this closed-loop tuning process until satisfactory values were reached.

In Figure 5(b-c) we show the spherical and cylindrical power measurements over the entire lens aperture, along with the calculated mean value and the std of each measurement. We reached a mean value of 4.025 $m^{-1}$ in spherical power, i.e. an error of 0.025 $m^{-1}$ relative to the nominal design, and a std of 0.268 $m^{-1}$. We obtained cylindrical power with an offset of 0.547 $m^{-1}$ from the nominal (C = 0) and a std of 0.271 $m^{-1}$. The area with the largest deviations from the nominal target powers is observed near the frame edges.

We attribute these deviations primarily to perturbations in the 3D printed frame, arising from the accuracy limits of our SLA printer, which translate into imperfect boundary conditions for the liquid lens. This has two consequences – localized deviations around the frame, and an overall violation of the cookie cutter conditions. i.e. while the neutral buoyancy conditions guarantee that the surface has constant mean curvature and thus there exists a volume for which the nominal spherical power can be retrieved, there is no volume that would retrieve the cylindrical power. This is supported by our computational results presented in SI.9, where we show the effect of perturbations on the frame's height and of deviations in its footprint shape on the cylinder power map, for a fixed nominal spherical power. For example, for a superposition of three localized deviations of 20 μm in the height of the rim, each spread over 10 mm, together with a planner deviation in the horizontal axis in a magnitude of 200 μm, the resulting deviation in the mean cylindrical power reaches to 0.308 $\mu m^{-1}$ with a standard deviation of 0.478 $\mu m^{-1}$, similar to the values obtained in the experiment presented in figure 5. A separate source of overall error in the shape of the surfaces is our manual injection process, which we estimate to have an accuracy of ~100 μL, exceeding the desired resolution of 26 μL (as discussed in Figure 4 and in SI.8). Lastly, under neutral buoyancy, geometrical deviations or volume inaccuracy should not affect the uniformity of the spherical power. Hence, the observed spatial variation in the spherical power must be the result of either deviations from neutral buoyancy conditions, or of shrinkage during polymerization.

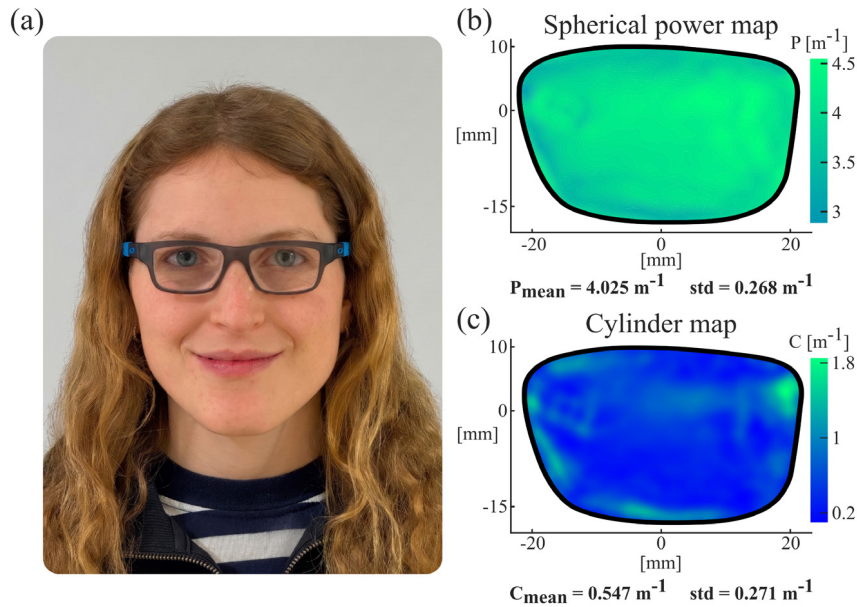

*Figure 5* – *Quality assessment of the fabricated lenses using the Cookie Cutter algorithm. (a) A photo illustrating the visual and cosmetic quality of the lenses in day-to-day usage, the lens appears clear and smooth up to the rim, with no visible defects near the edges. (b-c) Optical properties measurements of the left lens (as of the 'consumer' view). (b) Spherical power measurement of the solid lens throughout the entire lens aperture. The mean value of the spherical power is 4.025 $m^{-1}$, whereas the nominal target value is 4 $m^{-1}$, the std across the entire lens is 0.268 $m^{-1}$. (c) Cylindrical power measurement of the solid lens throughout the entire lens aperture. The mean value of the cylinder is 0.547 $m^{-1}$, whereas the nominal target value is 0 $m^{-1}$, the std across the entire lens is 0.271 $m^{-1}$. In both maps, the most significant deviations from the nominal values occur near the outer perimeter of the lens, stemming from the quality of the 3D printed frame, in the form of deformations in the topography of the frame's contours, and deviations in the volume injected into the frame from the nominal target.*

**DISCUSSION AND CONCLUSIONS**

The method we presented shows the ability to produce ophthalmic lenses via Fluidic Shaping directly on any desired rim shape, thus eliminating all mechanical processes, including edging. The algorithm we developed provides a straightforward method for computing the height variations along the rim-shaped frame and the required injection volume, that together yield the desired optical prescription. The computation can be readily implemented using standard CAD tools, and the computed frame can be quickly produced by 3D printing and be used as the boundary condition for Fluidic Shaping.

The work successfully demonstrates the feasibility and potential of this approach to produce eyeglasses without mechanical processes and without generating waste. However, it also highlights the challenges that must be overcome in order to reach high-quality lenses at industry standards. Perhaps the most significant challenge is achieving high accuracy of the 3D-printed frame, as imperfections in the frame geometry lead to surfaces deviations from the designed lens surfaces. A second challenge is precise injection of the liquid volume. As we showed, the required accuracy is in the order of tens of microliters over a nominal volume of 5-10 ml. While precise syringe pumps exist, even at the nanoliter scale, the time to complete an injection using a precision system would be extremely long. A more reasonable approach would be to inject the majority of the volume using a low accuracy but high speed pump, and then make fine-adjustments using a high accuracy one. These adjustments must however be performed in closed-loop to reach the desired optical prescription. The capsule we developed as part of this work shows the feasibility of making such measurements of the lens while still in liquid form, but our 'closed-loop' process was so far manual and iterative.

Lastly, as highlighted also by Elgarisi et al,[30,32] the path toward high-quality lenses produced with Fluidic Shaping passes through the identification or development of optical polymers with good mechanical and optical properties, and low shrinkage. It is very likely that two-component polymers would be advantageous over photopolymers, as they eliminate the dependence on uniform illumination and guarantee more uniform polymerization. However, the RTV we used in this work requires an excessively long time to polymerize in water (~24-48 hr), during which environmental changes may affect the buoyancy conditions, and some penetration of the immersion liquid into the polymer may occur. Furthermore, RTV is a relatively soft polymer and thus not ideal for ophthalmic application. Thus, hard polymers with a characteristic reaction time on the order of minutes – long enough to allow fine-tuning of the injection volume but short enough for practical applications – are highly desired. With such polymers, and further improvement in the accuracy of 3D printers, production of ophthalmic lenses without edging may become a reality.

**ACKNOWLEDGMENTS**

Funded by the European Union (ERC, Fluidic Shaping, 101044516). Views and opinions expressed are however those of the author(s) only and do not necessarily reflect those of the European Union or the European Research Council Executive Agency. Neither the European Union nor the granting authority can be held responsible for them.

We are very grateful to Shamir Optical Industry for supporting this work, and particularly for providing us with access to their metrology devices throughout this work, and for supplying the eyewear rims used in the experiments.

We thank Dr. Raanan Bavli, Dr. Asaf Szulc and Yiska Fatta from Rotlex for their advice and support in measurement of the lenses using the capsule device.

We thank Eden Shatzman for modeling the glasses we produced as part of this work.

Y.K. acknowledges the support and funding of the Irwin and Joan Jacobs Fellowship, and is grateful for the fellowship.

The authors acknowledge the use of Google Gemini 3.1 Pro (accessed March 2026) to modify the background color of Figure 5, as well as the Google Photos 'Portrait Light' feature for color enhancement (the original, unaltered image is provided as Figure S.10 in the Supplementary Information).

**AUTHOR CONTRIBUTIONS**

Y.K and M.E. performed the laboratory experiments. A.A.H and Y.K developed the theory and performed the numerical experiments. All authors contributed to analysis of the results. Y.K. and A.A.H. prepared the figures. Y.K. and M.B. wrote the manuscript. All authors reviewed and commented on the manuscript.

**DECLARATION OF COMPETING INTEREST**

The authors declare the following financial interests/personal relationships which may be considered as potential competing interests: Moran Bercovici reports financial support was provided by Technion Israel Institute of Technology. If there are other authors, they declare that they have no known competing financial interests or personal relationships that could have appeared to influence the work reported in this paper.

**DATA AVAILABILITY**

Data will be made available on request.

# Supplementary information

# Zero-waste manufacturing of ophthalmic lenses by direct fluidic shaping in arbitrary domains

Yotam Katzman[1], Mor Elgarisi[1], Amos A. Hari[1], Jonathan Ericson[1], Omer Luria[1], Valeri Frumkin[1,2], Moran Bercovici[1,*]

[1] Faculty of Mechanical Engineering, Technion – Israel Institute of Technology, Haifa, Israel
[2] Current affiliation: Department of Mechanical Engineering, Boston University, MA, USA
* Corresponding author: mberco@technion.ac.il

### SI.1 – Estimate of annual material waste in ophthalmic processing

- As mentioned in the introduction of the paper, the global eyewear industry produces around 2-3 million lenses each day[1–4].
- The initial semi-finished blank (SFB) weighs 50-70 grams (see table S.1 for details of several SFBs available commercially), and a typical final lens weighs around 2-5 grams.[5] Hence, in the surfacing process 90-97% of the material is discarded.
- Multiplying the number of annually produced lenses by a conservative SFB weight of 50 gr, and a conservative material loss of 90% per lens, yields 49,275 tons/year.

*Table S.1* - *Weight measurements of several commercial SFBs.*

| Manufacturer | Product details | Measured weight [gr] |
|---|---|---|
| Shamir | D 76 mm, Sag 1.19, TC 6.4, B 2, MR7, RI 1.67 | 66.04 |
| Shamir | D 81 mm, Sag 3.23, TC 5.39, B 5.5, Poly HC, RI 1.59 | 62.05 |
| Essilor | D 70 mm, CT 11, B 8 (Unifocal ORMA 15) | 57.73 |

### SI.2 – Estimation of indices of refraction for the immersion liquid and liquid polymer

Table S.2 provides the refractive indices of water with varying amount of glycerol as reported by R. Bochert et al.[6] In our case, the immersion liquid contained 27% glycerol in water. By linear interpolation, the refractive index of the immersion liquid is approximated to be 1.36.

*Table S.2* - *Refractive index as a function of glycerol content in water, as reported by R. Bochert et.al.*

| Percentage glycerol/distilled water [%] | Refractive index |
|---|---|
| 5 | 1.337 |
| 20 | 1.355 |
| 40 | 1.381 |
| 60 | 1.411 |
| 80 | 1.440 |
| 100 | 1.470 |

The manufacturer specifies a refractive index of 1.525 for the Vida Rosa in its solid state. This value is typically larger than the index of refraction in liquid state due to shrinkage, and can thus be regarded as an upper bound.[7] This is consistent also with indirect measurements of the index of refraction in liquid state described in SI.5, yielding refractive index of 1.453.

**SI.3 – Qualitative comparison between flat frame and varying edge-height frame**

Figure 2 presents a qualitative comparison between a flat frame and the varying edge-height frame, where we injected into both frames the same nominal volume, as derived by the Cookie Cutter algorithm. Here we show another comparison, in which we intentionally reduced the injected volume in the 'flat' frame such that the curvature of the liquid lens in the flat frame was, by side-view visual inspection, similar to that of the liquid lens formed in the varying edge-height frame (R ~ 129 mm). As shown in Figure S.3, the resulting lens remains distorted, further demonstrating that no nominal injection volume can yield the desired optical properties when using a flat frame geometry.

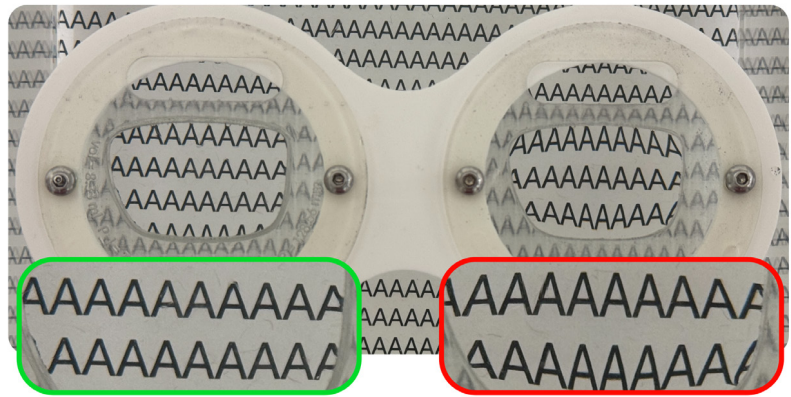

***Figure S.3*** *– Comparison of a varying edge-height frame (left, green) with the nominal liquid volume, and a flat frame (right, red) with a liquid volume that yields a radius of curvature similar to that of the desired spherical lens - R ~ 129 mm.*

**SI.4 – Capsule calibration**

Throughout the work, we used the Rotlex Mapper to measure the optical properties of the lenses while they were still in liquid form. To that end, we designed a sealed capsule (described in detail in the context of Figure 3 and 4 in the main text) that contains the liquid lens and fits within the imaging chamber of the Rotlex Mapper. However, the Mapper is calibrated for analysis of solid lenses in air, and thus any optical readings from lenses submerged within an immersion liquid must be corrected to represent their optical performance in air. If the Mapper's light source were collimated, then a naïve approach for correction would be to account for the differences in indices of refraction (air – polymer vs immersion liquid – polymer). However, the fact that the Mapper works using non-collimated light, together with the presence of the capsule's acrylic ceiling in the optical path, makes this correction non-trivial to model. We thus turn to calibration experiments to determine the transformation function.

Figure S.4 presents the spherical power of solid Vida Rosa lenses measured within the immersion liquid, as a function of the spherical power of the same lenses when measured in air. A total of nine lenses were tested, spanning a range of mean spherical powers from -3.5 to +5 diopters. To ensure consistency between the corresponding measurements, each lens was positioned in the exact same location and alignment during both measurement states. To that end, we marked a small dot at the center of each lens and a horizontal line at the top edge of the lens frame, and aligned them with the Mapper's crosshair and crossline, respectively. To further minimize angular-positioning related errors between paired measurements, we used only a small, 5×5 mm region at the center of each lens.

By performing a linear fit across all measurements, we derived a transfer function that enables prediction of the optical properties of the submerged lenses as if they were outside the capsule (in air),

$$P_a = \alpha \cdot P_w + \beta, \tag{1}$$

where $P_a$ is the spherical power measured in air, $P_w$ is the corresponding measurement inside the capsule filled with immersion liquid, and $\alpha$ and $\beta$ are the fitting coefficients.

We repeated the process for PDMS lenses using three lenses with diopters between $-3$ and $+4$, and we present this in Figure S.4. Since in both cases the polymer's refractive index and that of the immersion liquid vary between the two materials (due to differences in material properties and the adjusted glycerol-water ratio), each polymer resulted in a different transformation function, as summarized in Table S.3. The coefficient $\alpha$ represents the expected effect of the change in refractive index on the power. $\beta$ represents an additional constant bias introduced by the capsule and the liquid medium itself, acting as a weak optical element even in the absence of a lens (due to the fact that the Mapper's light source is non-collimated).

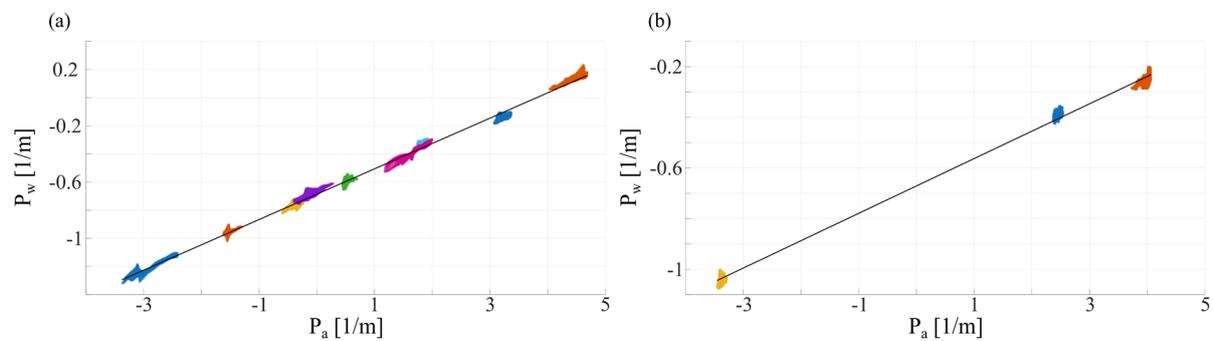

*Figure S.4 - Spherical power of (a) nine solid Vida Rosa lenses, and (b) three solid PDMS lenses measured in the capsule, as a function of their corresponding measurements in air. Each color corresponds to a different lens, with data points representing the measurement values within a 5×5 mm square at the center of each lens. The line in both plots corresponds to a linear fit to all data points.*

*Table S.3 - Transfer function parameters relating the spherical powers of solid lenses measured in the capsule to those measured in air, for Vida Rosa (n=1.525) and PDMS (n=1.41). The table presents the linear fit coefficients α and β, along with the corresponding RMSE and R² values, which indicate the accuracy and fit quality of each transformation.*

|  | α | β | RMSE m$^{-1}$ | R² |
|---|---|---|---|---|
| **Vida Rosa spherical power transfer function (based on nine solid lenses)** | 5.534 | 3.802 | 0.1244 | 0.9968 |
| **PDMS spherical power transfer function (based on three solid lenses)** | 8.922 | 6.2 | 0.2572 | 0.9931 |

Assuming that the transfer function remains linear also for other materials, the calibration process can be further simplified by taking only two data points. One data point must be constructed from measurement of a lens in air and in the liquid. The other data point may correspond to another lens with a different power, but further simplification can be achieved by measuring the power of a capsule without a lens, for which $P_a$ is known to be zero. We applied this method to calibrate the capsule for RTV-601, a two-component silicone polymer (n=1.41) and measured the data points [$P_a = 0$, $P_w = 0.67$] and [$P_a = 2$, $P_w = -0.46$], from which we obtain

$$P_a(RTV) = 9.91 \cdot P_w(RTV) + 6.64. \tag{2}$$

Because RTV-601 has very low shrinkage, it is reasonable to assume that its index of refraction does not change significantly between its liquid and solid states. Based on this assumption, the transfer function Equation 2 relates the power of *liquid* RTV-601 in the capsule to its expected *solid* power in air. Relying on this, we successfully fabricated RTV-601 lenses with the desired prescription, as presented in Figure 5.

### SI.5 - Estimating the refractive index of liquid Vida Rosa

Figure 4 in the main text shows the spherical power sensitivity to volume deviations for liquid lenses created on different frames. Producing the figure requires knowledge of the refractive index of Vida Rosa in its liquid form, which was not known *a priori*. We here detail our approach for estimating this value:

- Figure S.5.1 presents the computational results for the deviation of the mean curvature of the lenses from their nominal value, as a function of the deviations in the injection volume. These results are independent of the refractive index, and are a function of the computed geometry alone.

- Figure S.5.2 presents the measured spherical power of the liquid Vida Rosa lenses as a function of the deviations in the injected volume. Notably, these are raw measurements as obtained by the Mapper while the lenses were submerged in the immersion liquid within the capsule.

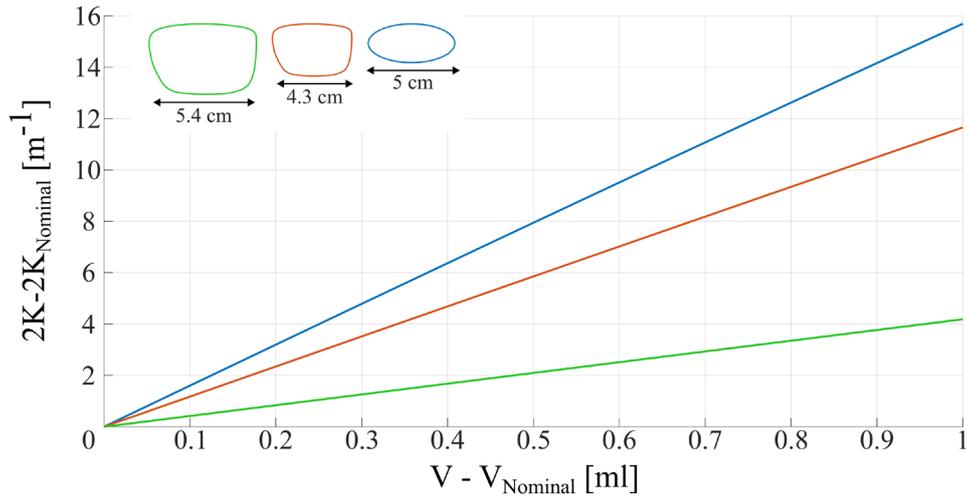

***Figure S.5.1*** – *Computational results presenting the deviation of the simulated lenses' double mean curvature for the three representative frames, as a function of their volume deviations. Each color corresponds to a different simulated lens.*

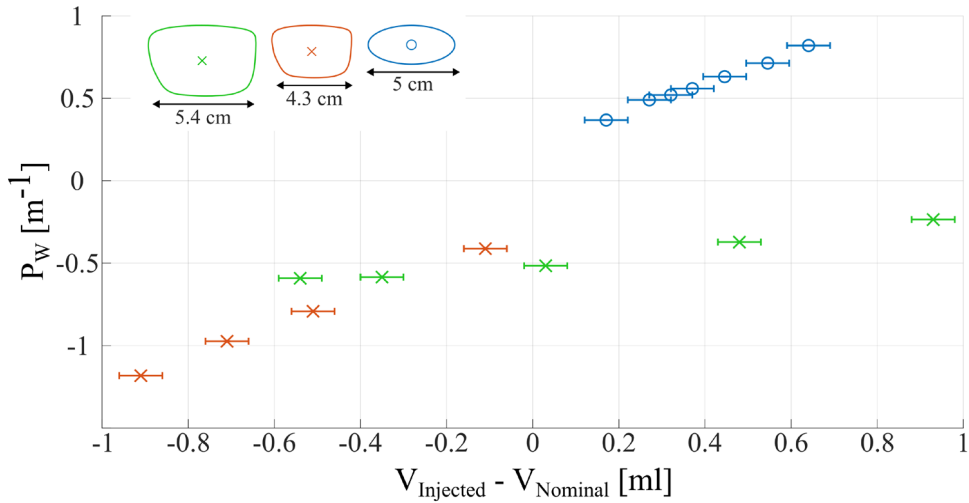

***Figure S.5.2*** – *Raw spherical power measurements of liquid lenses made of Vida Rosa, measured in the capsule filled with immersion liquid, as a function of their volume deviations. Each data point presents the average measured diopter value within a 3×3 mm square at the center of the lens. Each color corresponds to a different frame design. Horizontal error bars represent the uncertainty in injection volume (±50 μL).*

Clearly, both results show linear variation of the curvature or of the measured spherical power as a function of the deviation in volume, with sensitivity increasing as the frame size decreases. However, they use different metrics for their power (geometric power vs. optical power in the capsule). To directly compare them, we seek to transform both to optical power in air.

- The spherical power, $P_a$, of a bi-convex lens in air is related to its double curvature, 2K, through

$$P_a = (n_{VR,l} - 1) \cdot 2K, \qquad (3)$$

where $n_{VR,l}$ is the refractive index of the liquid polymer, which we seek to find.

- To transform the experimental spherical power measurement in the capsule to its equivalent power in air, we must subject it to the linear transformation: $P_a = \alpha \cdot P_w + \beta$, where $\alpha(n_{VR,l})$ and $\beta(n_{VR,l})$ are both also functions of (the yet unknown) $n_{VR,l}$. Figure S.4 presents the transformation functions for n=1.525 (solid Vida Rosa) and for n=1.41 (solid PDMS). Since the refractive index of liquid Vida Rosa is assumed to be slightly smaller than that of its solid state,[7] we expect its transformation function to be bounded by the functions of the solid PDMS and the solid Vida Rosa. Therefore, we can express the transfer function for the liquid state of Vida Rosa as a linear interpolation of these two known transfer functions,

$$P_a(n_{VR,l}) = \left[\alpha_1 + (\alpha_2 - \alpha_1) \cdot \frac{n_{VR,l} - n_1}{n_2 - n_1}\right] \cdot P_w + \left[\beta_1 + (\beta_2 - \beta_1) \cdot \frac{n_{VR,l} - n_1}{n_2 - n_1}\right], \quad (4)$$

where $\alpha$ and $\beta$ are the transfer function coefficients, and the subscript notations 1,2 represent PDMS and Vida Rosa in solid state, respectively (see table S.3 for their exact values).

If the computational results are correct, then there exists a value of $n_{VR,l}$ for which the computational (Equation 3) and the experimental (Equation 4) results collapse onto one another, for all frames. Using least mean-squares, we find that value to be $n_{VR,l} = 1.453$ with an $R^2$ value of 0.98, indicating excellent agreement between the two data sets, thus providing cross-validation for both the computation and the transformation. Figure S.5.3 presents the overlayed results of both the computations and the experiments using the obtained transformation function, $P_a = 7.64 \cdot P_w + 5.29$.

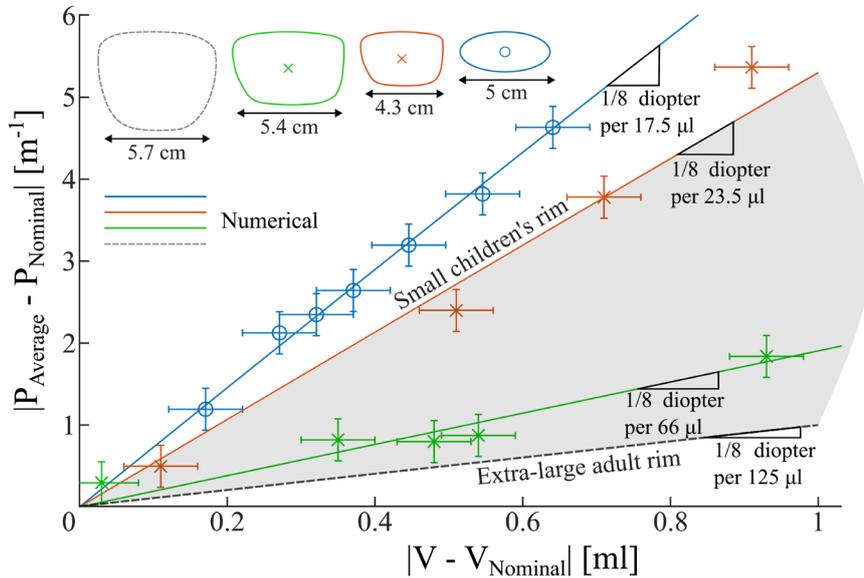

*Figure S.5.3 – Experimental and numerical results of spherical power sensitivity to volume variations for liquid Vida Rosa lenses, based on Equations 3, 4 using $n_{VR,l} = 1.453$. The excellent agreement between the transformed experimental measurements and the transformed numerical solution for all frames provides validation for both the numerical solution, and the transformations.*

## SI.6 – Raw data used for the creation of Figure 4

In Figure 4 we presented a sensitivity analysis of the spherical power to deviations in the injected volume. Table S.4 presents the raw data on which this figure is based. For each type of frame, we provide the nominal injection volume, nominal spherical power, the measured injection volume based on readings from the syringe and/or pipette, the average spherical power in a 3×3 mm square at the center of each lens, and the standard deviation for each such measurement.

*Table S.4 - Measured values used in Figure 4.*

| Lens frame type and number | Nominal volume [ml] | Injected volume [ml] | Nominal diopter [1/m] | Average diopter in a square of 3×3 mm [1/m] | Measurements std [1/m] |
|---|---|---|---|---|---|
| Elliptic #1 | 9.309 | 9.95 | 6.968 | 11.600 | 0.181 |
| Elliptic #2 |  | 9.48 |  | 8.160 | 0.033 |
| Elliptic #3 |  | 9.58 |  | 9.092 | 0.074 |
| Elliptic #4 |  | 9.63 |  | 9.316 | 0.041 |
| Elliptic #5 |  | 9.68 |  | 9.610 | 0.054 |
| Elliptic #6 |  | 9.75 |  | 10.163 | 0.116 |
| Elliptic #7 |  | 9.85 |  | 10.788 | 0.082 |
| Oakley #1 | 5.81 | 4.9 | 1.729 | -3.637 | 0.058 |
| Oakley #2 |  | 5.1 |  | -2.051 | 0.072 |
| Oakley #3 |  | 5.3 |  | -0.669 | 0.090 |
| Oakley #4 |  | 5.7 |  | 2.225 | 0.130 |
| Scaled-up Oakley #1 | 9.192 | 8.65 | 1.730 | 0.858 | 0.119 |
| Scaled-up Oakley #2 |  | 8.84 |  | 0.911 | 0.185 |
| Scaled-up Oakley #3 |  | 9.22 |  | 1.436 | 0.231 |
| Scaled-up Oakley #4 |  | 9.67 |  | 2.527 | 0.232 |
| Scaled-up Oakley #5 |  | 10.12 |  | 3.568 | 0.352 |

## SI.7 – Cylindrical power sensitivity to deviations in the injection volume

In the paper we presented computational results for the spherical power sensitivity to variations in the injection volume along with their experimental validation (Figure 4). Here, we show in addition computations of the cylindrical power sensitivity to volume deviation, for an index of refraction of 1.453 (corresponding to Vida Rosa in its liquid form). Figure S.7.1 presents the computed mean cylindrical power over the entire lens, for four different frames – an elliptical frame, an Oakley Marshal (kids) frame, a 25% scaled-up Oakley Marshal frame, and an Oakley Slender adult (large) frame. Similar to the spherical power, the cylindrical power also shows a linear dependence on the volume deviation, though its slope $\left(\frac{dC}{dV}\right)$ is approximately 50% of the spherical power one $\left(\frac{dP}{dV}\right)$. In Figure S.7.2 we present cylindrical power maps for three volume deviations ($\Delta V$ = +25, +50 and +100 μL). In all cases, as the volume deviation increases, the uniformity of the cylindrical map decreases and accordingly the deviation of the mean cylindrical power increases. Interestingly, while the mean cylindrical power of the elliptic frame is most sensitive to the volume deviation, its standard deviation remains small (on the order of 1E-3) even for the largest volume deviation.

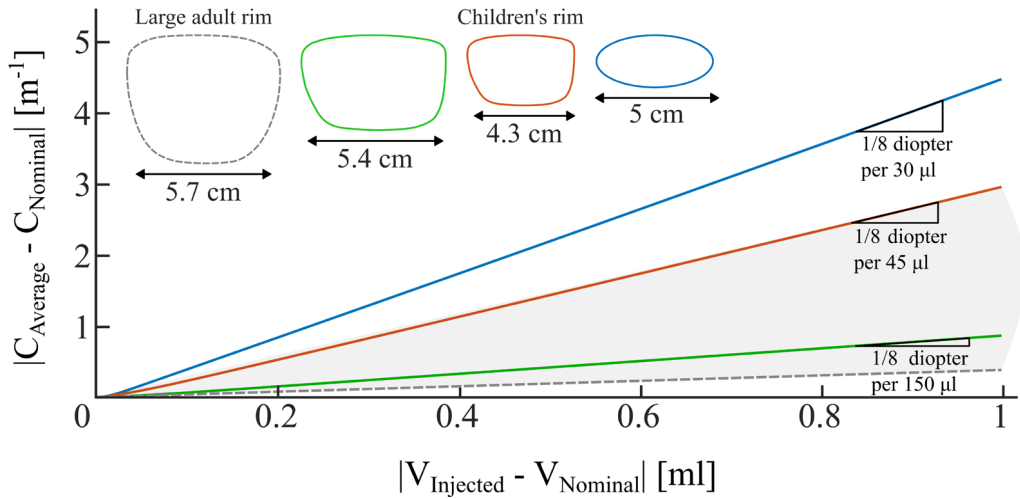

*Figure S.7.1 – Numerical results of cylindrical power sensitivity to volume variations for lenses formed in three representative frame designs. The plot shows the absolute deviation in cylindrical power (y axis) as a function of the absolute deviation in injected volume from its nominal value (x axis). Each line corresponds to a numerical solution for a different frame geometry. The four representative frames are: (1) a simple elliptical frame (blue), (2) a children's frame based on an Oakley eyewear rim (Oakley Marshal Xs OY 8005 Children's Glasses) (red) ,(3) a 25% scaled-up version of that same Oakley design (green), and (4) an Oakley Slender, representing a particularly large frame for adults (gray). The gray shaded area between the gray and red lines reflects the range of typical frame sizes. As derived from the slopes of the dashed lines, an injection accuracy in the range of ±45 (for very small frames) to 150 μL (for very large frames) is required to remain within the industry standard deviation of ±0.125 diopter.*

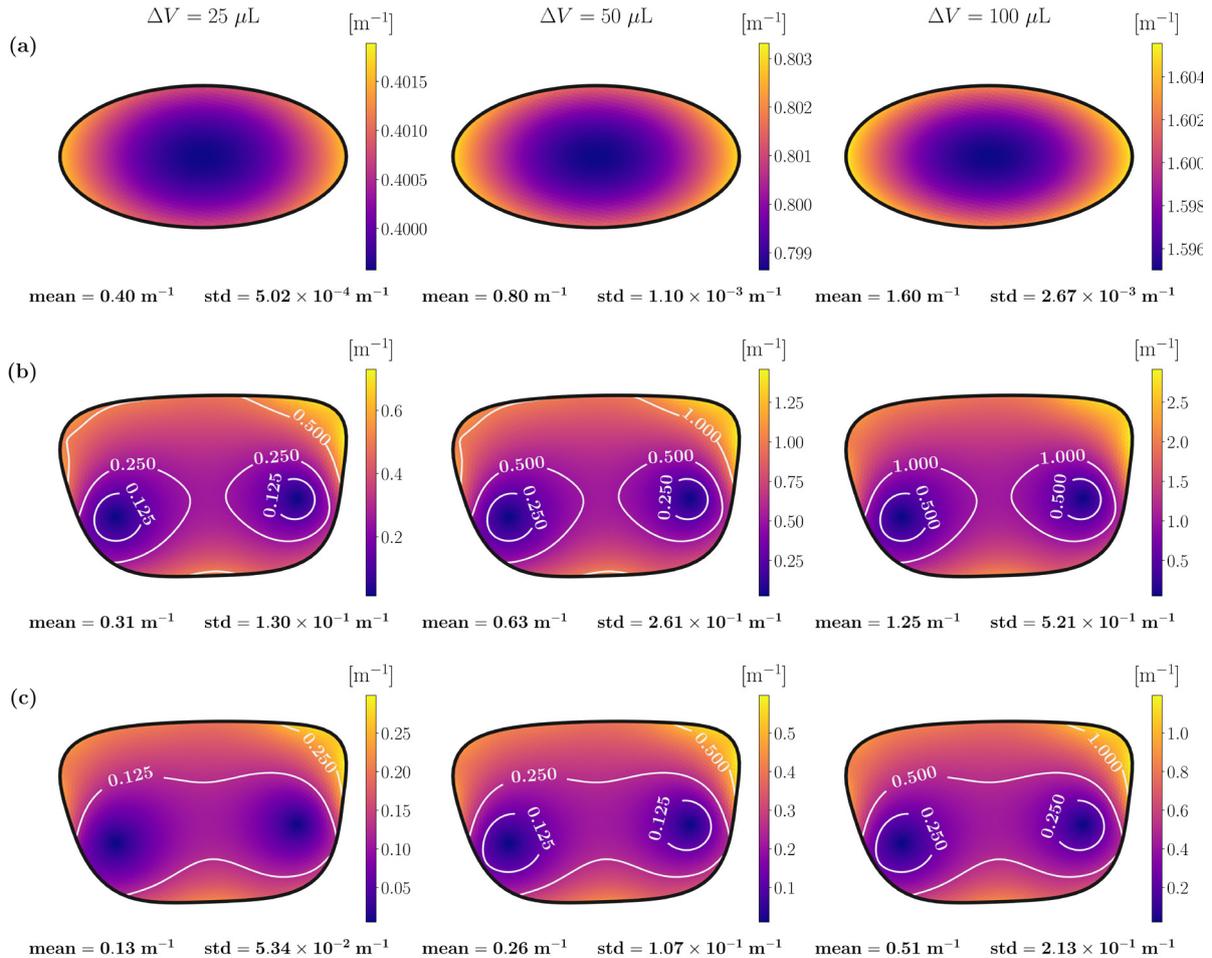

*Figure S.7.2* – *Simulated cylindrical power maps for three representative frames at various volume deviations. Each row represents a different frame: (a) elliptical frame, (b) Oakley Marshal frame, and (c) a 25% scaled-up version of the Oakley Marshal frame. Each column represents a different volume deviation: +25, +50, and +100 μL. Below each map, we present the calculated mean cylindrical power of the entire lens aperture, along with the corresponding standard deviation (std). As expected, larger deviations in the simulated injected volume lead to higher deviations from the nominal cylindrical power (C = 0) and increased std values. To guide the eye, we also show contour lines at powers of 1/8, 1/4, 1/2 and 1 m$^{-1}$. While the elliptical frame is most sensitive in its mean cylindrical power, its variation remains small and thus no contour lines appear. As expected, and is evident from b and c, a larger frame is less sensitive to the same deviation in volume.*

## SI.8 – Transformation of volume sensitivity requirements for various materials

From Figures 4 and S.7.1 we can deduce the required accuracy of injection needed to meet the ophthalmic industry standard. It is however important to note that the quantitative data presented in those figures is relevant only for the index of refraction of Vida Rosa polymer in its liquid state (n=1.453). To compute the sensitivity for other polymers, we must account for differences in index of refraction. We will here focus only on the spherical power, which is more sensitive (see section SI.7), but a similar approach can also be used for the cylindrical power. The lens makers equation relates the geometry to the power through

$$P = (n_{lens} - 1) \cdot \left(\frac{1}{R_{out}} - \frac{1}{R_{in}}\right) \tag{5}$$

where $n_{lens}$ is the refractive index of the lens, and $R_{in}, R_{out}$ are the radii of curvature of the inner and outer optical surfaces of the lens. Thus, for a given frame geometry, the power sensitivity to volume of one material can be related to another through,

$$\left(\frac{\Delta P}{\Delta V}\right)_2 = \frac{(n_1 - 1)}{(n_2 - 1)} \cdot \left(\frac{\Delta P}{\Delta V}\right)_1 \tag{6}$$

For example, the refractive index of RTV-601 is 1.41, as compared to 1.453 for liquid Vida Rosa in liquid state. Hence $\left(\frac{DP}{DV}\right)_{RTV} = \frac{1.453-1}{1.41-1} \cdot \left(\frac{DP}{DV}\right)_{VR}$. From Figure 4, for the Oakley Marshal frame $\left(\frac{DP}{DV}\right)_{VR} = 188 \frac{1}{m \cdot \mu L}$, and thus $\left(\frac{DP}{DV}\right)_{RTV} = 208 \frac{1}{m \cdot \mu L}$ for the same frame. Hence, the required injection accuracy for the RTV-601 is 26 $\mu L$.

**SI.9 – The effect of perturbations in frame geometry on the cylindrical power**

In this section we seek to characterize the effect of perturbations in the frame geometry on the cylindrical power of the lens. The results are used to (1) estimate whether the residual non uniformity in Figure 5 can be attributed to inaccuracies in our frame shape, which is at present printed using a consumer-grade 3D printer, and (2) inform the requirements for future frame fabrication to achieve a lens quality that meets the optical industry standard.

To that end, we utilize our computational tool and simulate two types of manufacturing defects that are likely to occur with 3D printing:

(1) A localized defect in the frame's height corresponding to inaccuracies in the printing process or defects in handling of the frame after printing. We examine the effect of the defect's location, amplitude, and width.
(2) A planar deviation in the frame's shape, corresponding to compression of the frame in one direction, as may occur due to non-planar positioning in the printer. We explore the effect of the magnitude of this compression.

As noted also in the main text, the constant-mean-curvature nature of the liquid interface under neutral buoyancy guarantees that the spherical power is uniform for any injected volume. In all of our numerical experiments we have thus set the injected volume such that the obtained spherical power is equal to the nominal one (for an unperturbed frame). We thus do not show the spherical power, and focus only on the effect of the perturbations on the cylindrical power.

Figure S.9.1 presents the effect of the position of a localized defect on the cylindrical power. The defect has a Gaussian-shape, with a constant amplitude of 20 μm and a constant width of 10 mm (corresponding to 3σ), and is positioned at different locations along the frame. In all cases, the deviation in cylindrical power is localized, and decays to 1/8 diopter approximately 1 cm into the lens. The influence on the cylindrical power is nearly independent of the position along the frame, as indicated both by the power maps and by the mean and std values.

Figure S.9.2 presents the effect of the defect width on the cylindrical power. Here, we again use a Gaussian defect with an amplitude of 20 μm, and varied its width between 2.5 mm and 20 mm. As the width increases, we see that on one hand the deviation in power in the vicinity of the defect decreases significantly, and on the other hand the penetration of the deviation (e.g. the 1/8 diopter line) extends further into the center of the lens. This is consistent with the theory for a circular domain in Elgarisi et al., which showed sinusoidal variations in the height of the frame decay over shorter distances as the variation's wavelength increases.

Figure S.9.3 presents the effect of the defect amplitude on the cylindrical power. We used the same Gaussian-shaped defect as in S.9.1 and S.9.2, but here we fixed the width at 10 mm, and varied its amplitude between 20 μm to 1000 μm. Clearly, the power is extremely sensitive to the amplitude of the defect, where even 20 μm defects already create unacceptable deviations in the power. Single micron accuracy would thus be required to maintain an acceptable deviation - this is currently not feasible with consumer-grade printers, but is possible with high-end printers.

Lastly, Figure S.9.4 presents the impact of planar boundary deformations on the cylindrical power. Here, a horizontal planar compression ("squeeze") ranging from 20 μm to 1000 μm is applied to the lens boundary. The resulting cylindrical deviations are primarily concentrated at the central region of the lens. As the compression magnitude reaches 500 μm, the cylindrical power at the center of the lens (the most important optical zone) surpasses 1/8 diopters. Despite this central degradation, when compared to the localized defects analyzed previously, macroscopic planar deformations have a milder influence on overall optical performance. Notably, a localized boundary defect with an amplitude of 20 μm induces a disruption to the central optical zone comparable to a planar squeeze of 500 μm. We thus conclude that any future choice of manufacturing technology for the frame should emphasize accuracy of out-of-plane deformations (i.e. frame height) over in-plane ones.

In Figure S.9.5 we attempt to approximate the deviations obtained in the lens produced in our experiments (Figure 5), using a combination of compression and local defects. We note that we have not applied rigorous optimization to set the deformations locations and magnitude to match the measured power map precisely. Nevertheless, by trial and error and visual assessment, we can estimate that the experimental deviations may result from three main local deformations with magnitude on the order of 20 um over a 10 mm width, and an overall compression with a magnitude of 200 um. These values are within the expected tolerance of the 3D printer we used. For completion, we also show the spherical power for this case, showing that, despite the multiple defects, it is indeed uniform - as expected from theory.

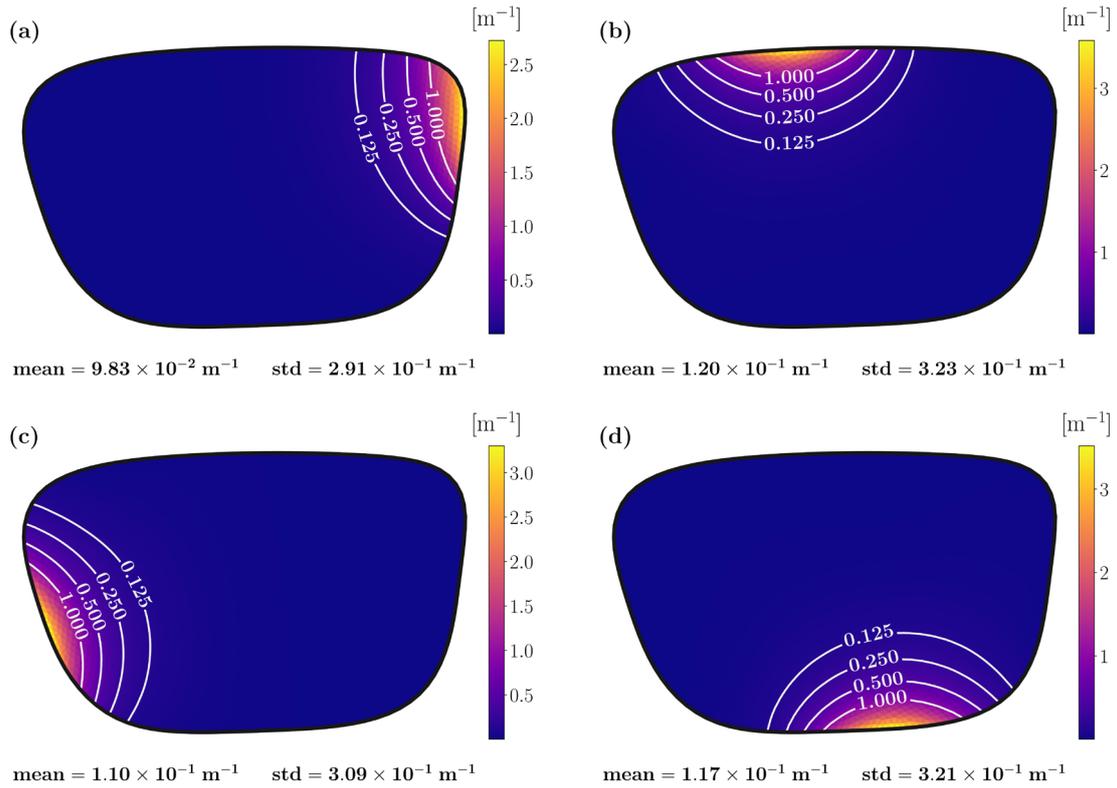

*Figure S.9.1* – *The effect of the position of a localized perturbation on the cylindrical power. Each simulation introduces a single Gaussian-shaped perturbation at a different position, all with an amplitude of 20 μm and a width of 10 mm. For each case, we adjust the liquid volume relative to the nominal value of 5.3 ml, such that the spherical power (not shown) remains fixed at 4 diopters. (a) ΔV = 0.95 μL. (b) ΔV = 1.57 μL. (c) ΔV = 1.28 μL. (d) ΔV = 1.50 μL.*

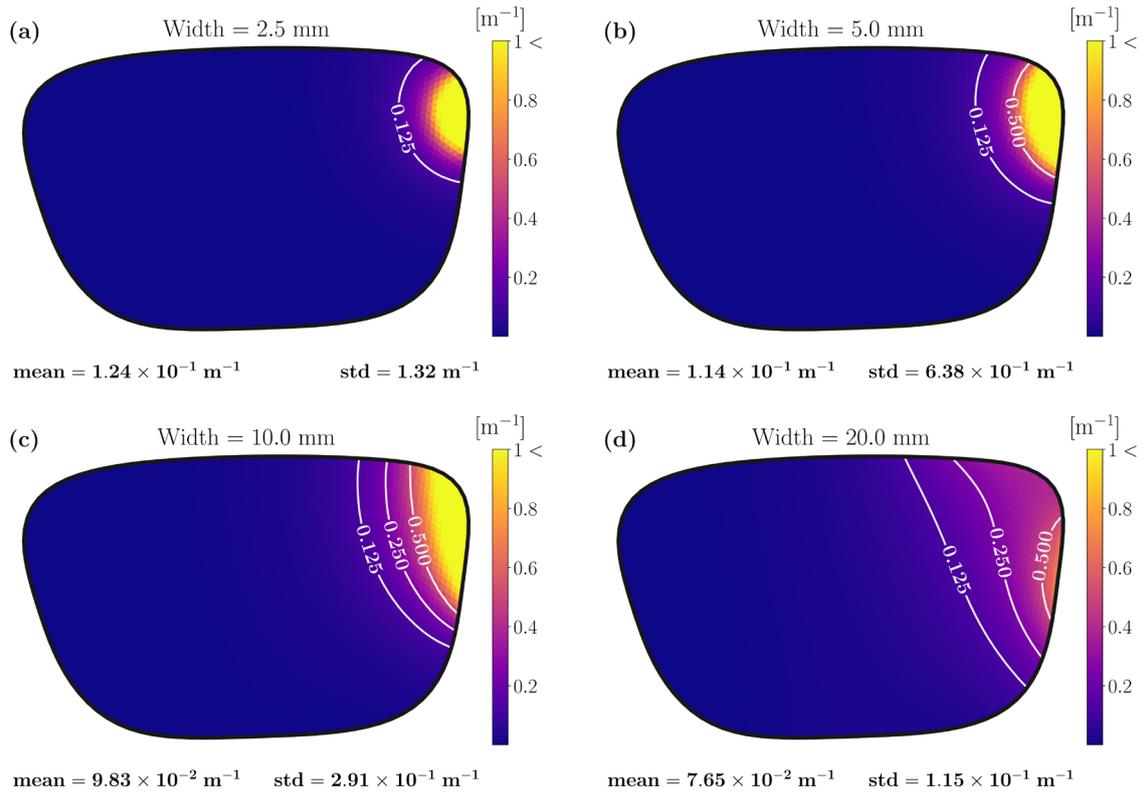

*Figure S.9.2* – *The effect of the width of a localized perturbation on the cylindrical power. Each simulation introduces a single Gaussian-shaped perturbation at the same location, with an amplitude of 20 mm, where the only changes between the simulations are the values of the perturbation's width and the corresponding adjusted volume. (a) Width = 2.5 mm, ΔV = 0.23 µL. (b) Width = 5 mm, ΔV = 0.46 µL. (c) Width = 10 mm, ΔV = 0.95 µL. (d) Width = 20 mm, ΔV = 2.02 µL.*

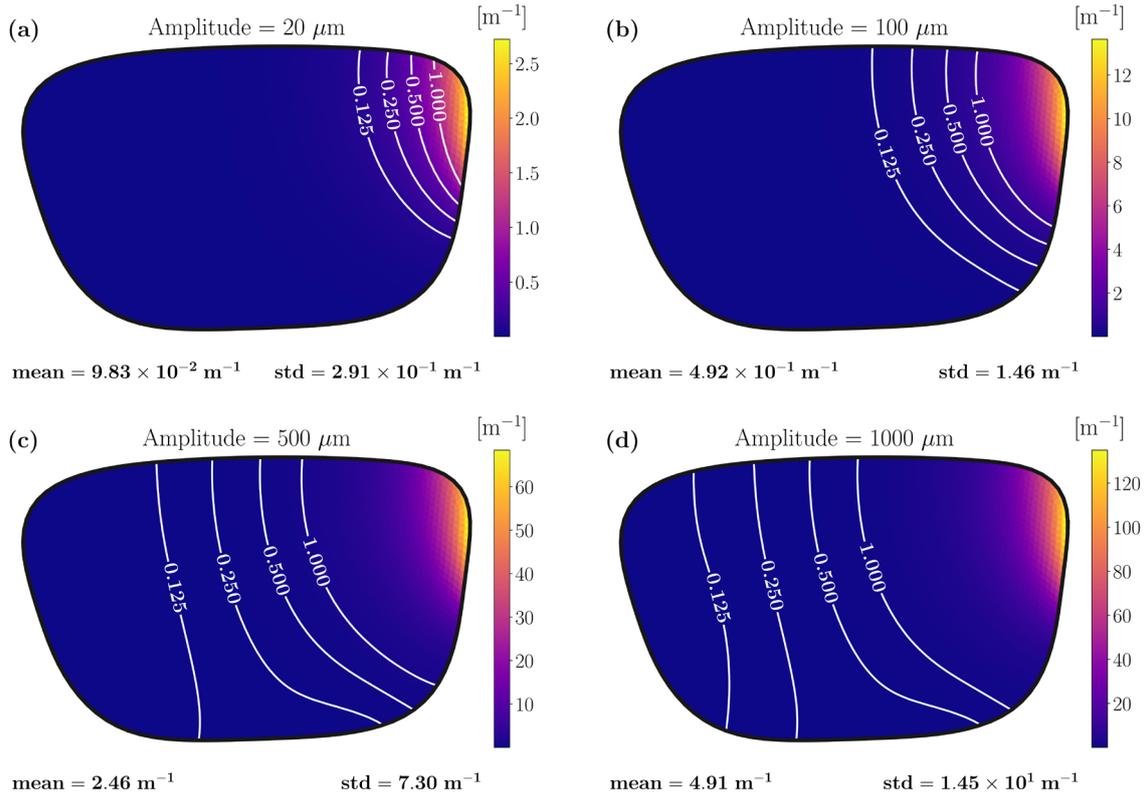

***Figure S.9.3*** – *The effect of the amplitude on the cylindrical power. Each simulation introduces a single Gaussian-shaped perturbation at the same location, with a width of 10 mm, where the only changes between the simulations are the values of the perturbation's amplitude and the corresponding adjusted volume. (a) Amplitude = 20 µm, ΔV = 0.95 µL. (b) Amplitude = 100 µm, ΔV = 4.74 µL. (c) Amplitude = 500 µm, ΔV = 23.76 µL. (d) Amplitude = 1000 µm, ΔV = 47.61 µL.*

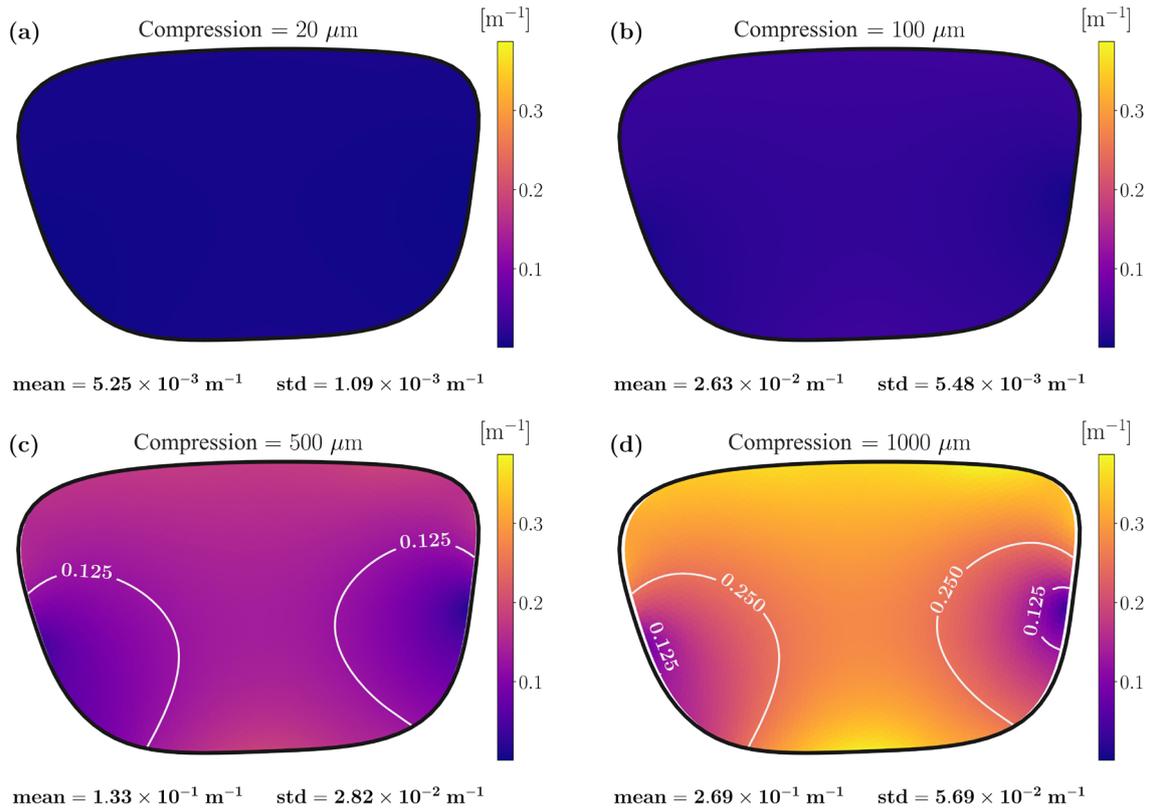

***Figure S.9.4*** *– The effect of the planar-deviation magnitude on the cylindrical power. Each simulation introduces a planar perturbation in the form of compression in the horizontal direction. The only changes between the simulations are the values of the magnitude of the perturbations and the corresponding adjusted volume. In all cases, the black frame corresponds to the original (uncompressed) geometry of the frame, whereas the colormap is bounded within the compressed shape. The difference is particularly evident in the 1000 um case. (a) Compression magnitude = 20 μm, ΔV = -2.13 μL. (b) Compression magnitude = 100 μm, ΔV = -10.65 μL. (c) Compression magnitude = 500 μm, ΔV = -53.27 μL. (d) Compression magnitude = 1000 μm, ΔV = -106.41 μL.*

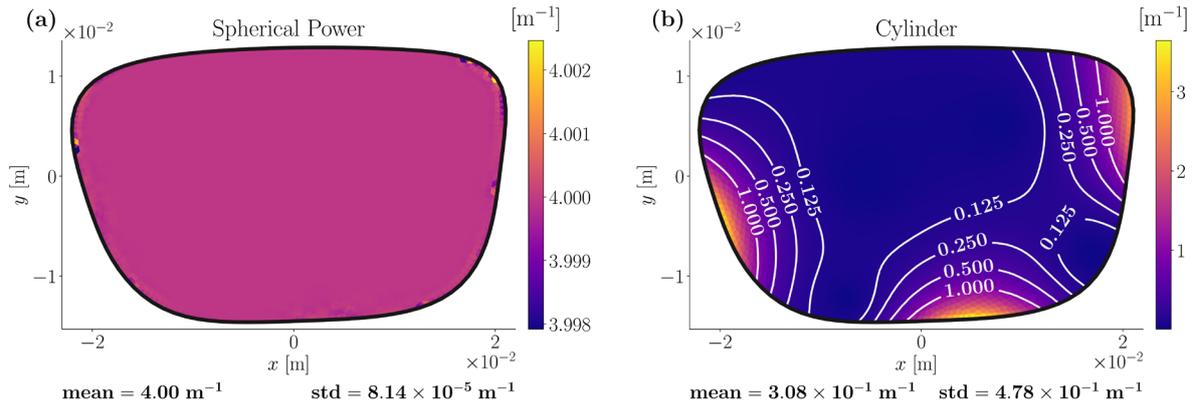

***Figure S.9.5*** *– Numerical results demonstrating the effect of multiple defects on the optical properties of the lens. Here we introduced a superposition of 3 local Gaussian perturbations to the height of the frame, all with an amplitude of 20 μm and a width of 10 mm, together with a planar compression perturbation with a magnitude of 200 μm. As in all previous cases, we adjusted the volume to match the spherical power to the nominal one, with ΔV = −17.6 μL. (a) As expected from theory, despite the multiple defects, the spherical power remains uniform. (b) The resulting cylindrical power map, showing a mean power of 3.08 x $10^{-1}$ $m^{-1}$, and the std is 4.78 x $10^{-1}$ $m^{-1}$. These values are similar to the values obtained in the experiment presented in Figure 5 (mean = 5.47 x $10^{-1}$ $m^{-1}$, std = 2.71 x $10^{-1}$ $m^{-1}$), suggesting that the deviations in the experiments may be explained by similar geometrical defects in the frame.*

### SI.10 – Figure 5 original image

We here present the original image (i.e. before editing as discussed in the acknowledgement section) from Figure 5a.

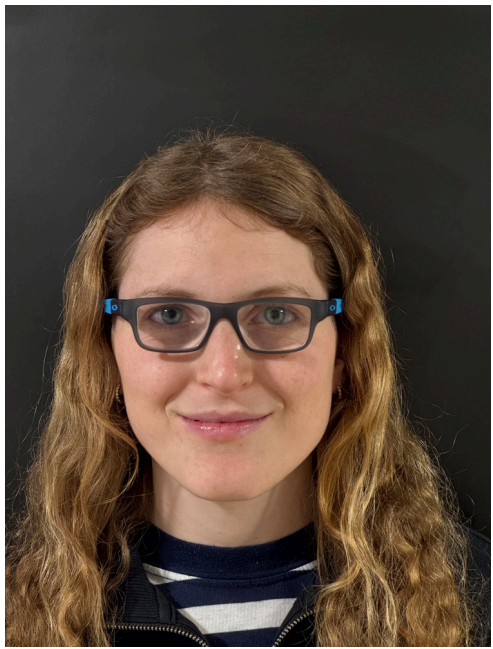

***Figure S.10*** *– the original image used in Figure 5.*